\documentclass[12pt]{article}
\usepackage{amssymb}
\usepackage{amsfonts}
\usepackage{amsmath}

\begin{document}

\begin{center}
{\Large A NOTE ON CALCULATING AUTOCOVARIANCES OF PERIODIC \textit{ARMA}
MODELS }

\ \ \ \ \ \ \ \ \ \ \ \ \ \ \ \ \ \ \

Abdelhakim A\textsc{knouche}

Hac\`{e}ne B\textsc{elbachir}

Fay\c{c}al H\textsc{amdi}
\end{center}

\ \ \ \ \ \ \ \ \ \ \ \ \ \ \ \ \ \ \ \ \

\noindent \textbf{ABSTRACT}

An analytically simple and tractable formula for the start-up
autocovariances of periodic $ARMA$ ($PARMA$) models is provided.

\ \ \ \ \ \ \ \ \ \ \ \ \ \ \ \ \ \ \ \ \



\ \ \ \ \ \ \ \ \ \ \ \ \ \ \ \ \ \ \ \ \ \ \ \ \

\noindent {\large 1. INTRODUCTION}

Autocovariance calculation procedures for $PARMA$ models are generally
carried out using the periodic Yule Walker equations (e.g. Bentarzi and
Aknouche, $2005$). This approach has been considered earlier by Li and Hui $%
(1988)$ for calculating $PARMA$ autocovariances, where the $(p+1)$-start-up
autocovariances, for all seasons, were given through a matrix equation $%
A\gamma =y$, which is solved for $\gamma $ ($\gamma $ being the $%
(p+1)S\times 1$-vector of the start-up autocovariances). The latter equation
is however analytically and computationally intractable since the matrix $A$
is not given explicitly but formed through an appropriate algorithm.
Adopting the same approach, Shao and Lund $(2004)$ showed that the $r$%
-start-up ($r=\max (p,q)+1$) autocovariances may be obtained by solving a
linear system $\Gamma U\gamma =\kappa $ for $\gamma $, where $\Gamma $ and $U
$ are matrices of dimensions $rS\times (p+1)rS$ and $(p+1)rS\times rS$,
respectively. While these matrices are given explicitly, the method remains
relatively cumbersome since it requires an increasing bookkeeping due to the
matrix product. This note proposes an improved computation procedure for
calculating the $PARMA$ autocovariances based on the latter approach. The
proposed method computes the $(p+1)$-start-up autocovariances based on a
linear system with a corresponding matrix given explicitly, whose analytical
form exhibits a circular structure, naturally assorted with the model
periodicity.

\ \ \ \ \ \ \ \ \ \ \ \ \ \ \ \ \ \ \ \

\noindent {\large 2. THE METHOD}

Consider a causal $PARMA$ model of orders $(p,q)$ and period $S$%
\begin{equation}
\overset{p}{\underset{j=0}{\sum }}\phi _{j}^{\left( v\right) }y_{v-j+nS}=%
\overset{q}{\underset{j=0}{\sum }}\theta _{j}^{\left( v\right) }\varepsilon
_{v-j+nS}\text{, }1\leq v\leq S\text{, }n\in \mathbb{Z},  \tag{$1$}
\end{equation}%
where $\phi _{0}^{\left( v\right) }=\theta _{0}^{\left( v\right) }=-1$ and $%
\{\varepsilon _{t},t\in Z\}$ is a periodic white noise process, i.e., a
sequence of uncorrelated random variables with mean zero and variance $%
E(\varepsilon _{v+nS}^{2})=\sigma _{v}^{2}$,$\ $for $1\leq v\leq S$ and $%
n\in \mathbb{Z}$.

Let $\gamma _{h}^{\left( v\right) }=E\left( y_{v+nS}y_{v+nS-h}\right) $ be
the autocovariance function at season $v$ and lag $h\in \mathbb{Z}$. Then,
it is well known (Li and Hui, $1988$; Shao and Lund, $2004$) that
multiplying $(1)$ by $y_{v+nS-h}$ and tacking expectation, the $\left(
\gamma _{h}^{\left( v\right) }\right) $ are completely identified from the
difference equation%
\begin{equation}
\gamma _{h}^{\left( v\right) }-\overset{p}{\underset{j=1}{\sum }}\phi
_{j}^{\left( v\right) }\gamma _{h-j}^{\left( v-j\right) }=-\overset{q}{%
\underset{j=h}{\sum }}\theta _{j}^{\left( v\right) }\psi _{j-h}^{\left(
v-h\right) }\sigma _{v-j}^{2}\mathbf{1}_{[h\leq q]},\text{ }h\geq 0,
\tag{$2$}
\end{equation}%
where the normalized cross-autocovariances $(\psi _{k}^{\left( v\right) })$,
coefficients of the unique causal representation of the $PARMA$ process $%
\{y_{t},t\in Z\}$, are given by (see e.g. Lund and Basawa, $2000$; Shao and
Lund, $2004$)%
\begin{equation}
\psi _{k}^{\left( v\right) }=-\theta _{k}^{\left( v\right) }\mathbf{1}%
_{[k\leq q]}+\sum\limits_{j=1}^{\min (k,p)}\phi _{j}^{\left( v\right) }\psi
_{k-j}^{\left( v-j\right) },\text{ \ \ \ }k\geq 1,\text{ }{\small v=1,...,S,}
\tag{$3$}
\end{equation}%
with $\psi _{0}^{\left( v\right) }=1$ ($\mathbf{1}_{\left[ .\right] }$\
stands for the indicator function).

Equation $(2)$ needs to be started from the knowledge of $\gamma
_{h}^{\left( v\right) }$, $0\leq h\leq p$ and $1\leq v\leq S$. Once these
start-up values are given, the $\gamma _{h}^{\left( v\right) }$ for $h>p$
may be obtained recursively from $(2)$ while invoking $(3)$. For $\gamma
_{h}^{\left( v\right) }$ with negative lags, we may use the well-known
relation $\gamma _{-h}^{\left( v\right) }=\gamma _{h}^{\left( v+h\right) }$.
The main result of this note is to formulate a linear system for computing
the $p+1$ necessary starting autocovariances. Let $\mathbf{\gamma }=(\gamma
_{0}^{\left( 1\right) },...,\gamma _{p}^{\left( 1\right) },\gamma
_{0}^{\left( 2\right) },...,\gamma _{p}^{\left( 2\right) },...,\gamma
_{0,}^{\left( S\right) },...,\gamma _{p}^{\left( S\right) })^{\prime }$ be
the $S(p+1)$-vector of such values and $\mathbf{\zeta }$ be the $S(p+1)$%
-vector whose entries $\mathbf{\zeta }_{hS+v}=\overset{q}{\underset{j=h}{%
\sum }}\theta _{j}^{\left( v\right) }\psi _{j-h}^{\left( v-h\right) }\sigma
_{v-j}^{2}$, for $1\leq v\leq S$ and $0\leq h\leq p$, are the right-hand
sides of $(2)$.

Define the $\left( p+1\right) $-square matrices $\mathbf{\varphi }_{h}^{(v)}$
($h\geq 0$, $1\leq v\leq S$) and $\mathbf{\Phi }_{k}^{(v)}$ ($k$, $v\in
\left\{ 1,...,S\right\} $) as follows%
\begin{equation}
\mathbf{\varphi }_{h}^{(v)}=\left\{
\begin{array}{l}
\left(
\begin{tabular}{l|l}
$\mathbf{0}_{h\times \left( p+1-h\right) }$ & $\mathbf{0}_{h\times h}$ \\
\hline
$%
\begin{array}{ccccc}
-\phi _{h}^{\left( v\right) } & -\phi _{h+1}^{\left( v\right) } & \cdots  &
& -\phi _{p}^{\left( v\right) } \\
0 & -\phi _{h}^{\left( v\right) } & 0 & \cdots  & 0 \\
& \ddots  & \ddots  & \ddots  & \vdots  \\
\vdots  &  &  & \ddots  & 0 \\
0 & \cdots  &  & 0 & -\phi _{h}^{\left( v\right) }%
\end{array}%
$ & $\mathbf{0}_{\left( p+1-h\right) \times h}$%
\end{tabular}%
\right) ,\text{ for }h=0,...,p \\
\mathbf{0}_{\left( p+1\right) \times \left( p+1\right) }\text{ \ \ \ \ \ \ \
\ \ \ \ \ \ \ \ \ \ \ \ \ \ \ \ \ \ \ \ \ \ \ \ \ \ \ \ \ \ \ \ \ \ \ \ \ \
\ \ \ \ \ \ \ \ \ \ \ \ for }h\geq p+1,%
\end{array}%
\right.   \tag{$4$}
\end{equation}%
\begin{equation}
\mathbf{\Phi }_{k}^{(v)}=\sum_{n\geq 0}\mathbf{\varphi }_{nS+k}^{(v)},\text{
}v,k\in \left\{ 1,...,S\right\} ,  \tag{$5$}
\end{equation}%
and the $S(p+1)$-square matrix%
\begin{equation*}
\mathbf{\Phi }=\left(
\begin{array}{ccccc}
\mathbf{\Phi }_{0}^{(1)} & \mathbf{\Phi }_{S-1}^{(1)} & \cdots  & \mathbf{%
\Phi }_{2}^{(1)} & \mathbf{\Phi }_{1}^{(1)} \\
\mathbf{\Phi }_{1}^{(2)} & \mathbf{\Phi }_{0}^{(2)} & \cdots  & \mathbf{\Phi
}_{3}^{(2)} & \mathbf{\Phi }_{2}^{(2)} \\
\vdots  & \vdots  & \ddots  & \vdots  & \vdots  \\
\mathbf{\Phi }_{S-2}^{(S-1)} & \mathbf{\Phi }_{S-3}^{(S-1)} & \cdots  &
\mathbf{\Phi }_{0}^{(S-1)} & \mathbf{\Phi }_{S-1}^{(S-1)} \\
\mathbf{\Phi }_{S-1}^{(S)} & \mathbf{\Phi }_{S-2}^{(S)} & \cdots  & \mathbf{%
\Phi }_{1}^{(S)} & \mathbf{\Phi }_{0}^{(S)}%
\end{array}%
\right) .
\end{equation*}%
where $\mathbf{0}_{m\times n}$ denotes the null matrix of dimension $m\times
n$. Then, the starting autocovariance vector $\mathbf{\gamma }$ is the
unique solution of the following linear system
\begin{equation}
\mathbf{\Phi \gamma =\zeta },  \tag{$6$}
\end{equation}%
whenever model $(1)$ is causal. Note that, in view of $(4)$, the infinite
sum in $(5)$ contains only $p$ non zero terms. It may be possible to reduce
the complexity of forming $\mathbf{\Phi }$ using its circular property.
Indeed, equation $(5)$ may be used to only evaluate the first bloc $\mathbf{%
\Phi }_{k}^{(1)}$, $k=1,...,S$. The blocs $\mathbf{\Phi }_{k}^{(v)}$ ($%
v=2,...,S$) would be deduced from $\mathbf{\Phi }_{k}^{(1)}$ by substituting
the corresponding parameters $\phi _{j}^{\left( 1\right) }$ by $\phi
_{j}^{\left( v\right) }$ for $j=1,...,p$.

\ \ \ \ \ \ \ \ \ \ \ \ \ \ \ \ \ \ \

\noindent {\large 3. CONCLUDING REMARKS}

Despite the simplicity of the proposed method, it has the drawbacks that the
starting autocovariances for all seasons are computed in the same bloc,
thereby requiring $O\left( (S(p+1))^{3}\right) $ operations, which might be
very costly for models with a large period. This is the main limitation of
the periodic Yule Walker approach compared to which the methods that compute
autocovariances for distinct seasons separately (e.g. Aknouche, $2007$) are
more suitable.

\ \ \ \ \ \ \ \ \ \ \ \ \ \ \ \ \ \ \ \ \ \ \ \ \ \ \ \

\noindent BIBLIOGRAPHY

\noindent Aknouche, A., $(2007)$. Causality conditions and autocovariance
calculations in $PVAR$ models. \textit{Journal of Statistical Computation
and Simulation}, \textbf{77}, 769--780.

\noindent Bentarzi, M. and Aknouche A. $(2005)$. Calculation of the Fisher
information matrix for periodic $ARMA$ models. \textit{Communications in
Statistics-Theory and Methods}, \textbf{34}, 891-903.

\noindent Li, W. K and Hui, Y. V. $(1988)$. An Algorithm for the exact
likelihood of periodic autoregressive moving average models. \textit{%
Communication in Statistics- Simulation and Computation}, \textbf{16},
1483-1494.

\noindent Lund, R. and Basawa, I. V. $(2000)$. Recursive prediction and
likelihood evaluation for periodic $ARMA$ models. \textit{Journal of Time
Series Analysis}, \textbf{21}, 75-93.

\noindent Shao, Q. and Lund, R. $(2004)$. Computation and characterization
of autocorrelations and partial autocorrelations in periodic $ARMA$ models.
\textit{Journal of Time Series Analysis}, \textbf{25}, 359-372.

\bigskip

Faculty of Mathematics/University of Sciences and Technology Houari
Boumediene,

BP 32,  El Alia, 16111, Bab Ezzouar,  Algiers, Algeria

aknouche\_ab@yahoo.com,

hacenebelbachir@gmail.com,

hamdi\_fay@yahoo.fr

\end{document}